\documentclass[prd,twocolumn,showpacs,amsmath,amssymb,superscriptaddress,preprintnumbers,nofootinbib,showkeys]{revtex4}

\begin{document}

\newcommand{\nablab}{{\mathop {\rule{0pt}{0pt}{\nabla}}\limits^{\bot}}\rule{0pt}{0pt}}

\title{The extended  Einstein-Maxwell-aether-axion model: \\ Exact solutions for axionically controlled pp-wave aether modes}

\author{Alexander B. Balakin}
\email{Alexander.Balakin@kpfu.ru} \affiliation{Department of
General Relativity and Gravitation, Institute of Physics, Kazan
Federal University, Kremlevskaya str. 18, Kazan 420008,
Russia}

\date{\today}

\begin{abstract}
The extended Einstein-Maxwell-aether-axion model describes internal interactions inside the system, which contains gravitational, electromagnetic fields, the dynamic unit vector field describing the velocity of an aether, and the pseudoscalar field associated with the axionic dark matter. The specific feature of this model is that the axion field controls the dynamics of the aether through the guiding functions incorporated into the Jacobson's constitutive tensor. Depending on the state of the axion field these guiding functions can control and switch on or switch off the influence of acceleration, shear, vorticity and expansion of the aether flow on the state of physical system as a whole.
We obtain new  exact solutions,  which possess the pp-wave symmetry, and indicate them by the term pp-wave aether modes in contrast to the pure pp-waves, which can not propagate in this field conglomerate.
These exact solutions describe a specific dynamic state of the pseudoscalar field, which corresponds to one of the minima of the axion potential, and switches off the influence of shear and expansion of the aether flow; the model does not impose restrictions on the Jacobson's coupling constants and on the axion mass.
Properties of these new exact solutions are discussed.

\end{abstract}
\maketitle

\section{Introduction}

The question: whether the velocity of gravitational wave propagation coincides with the speed of light, arose again as the result of extraordinary event, which enters the history of astronomy with the indication  GW170817/GRB 170817A. The date 17.08.2017 is associated now with the fact, that the gamma-ray burst GRB 170817A, observed by the Fermi Gamma-ray Burst Monitor, and the INTEGRAL (International Gamma-Ray Astrophysics Laboratory) from a binary neutron star merger, and  the gravitational-wave event GW170817, observed by the Advanced LIGO and Virgo detectors, had the time delay of $+1.74±0.05  \ {\rm s}$ \cite{170817}. The authors of the report \cite{170817} confirm that the difference between the speed of gravity and the speed of light to be between $-3\times10^{-15}$ and $+7\times 10^{-16}$ in terms of the speed of light. In other words, this outstanding discovery returns us to the idea that the mechanism of the gravitational wave propagation can be, in principle, superluminal and/or subluminal. In this context, we can consider also the possibility that the speed of gravity can be mediated by the influence of other fields and cosmic substrata (e.g., dark energy and/or dark matter), within which the gravity waves propagate. From this point of view, the cosmic dark fluid can control the speed of propagation of the gravitational waves, and we have to discuss examples of models describing such guidance. In this letter we consider an exactly integrable model, in which the dynamic aether and axionic dark matter interact with gravity waves. Generally, the existence of the dynamic aether and/or of the  massive pseudoscalar field prevents the gravity wave to propagate with the velocity equal to the speed of light in vacuum. However, below, based on exact solutions of model master equations, we have shown that this is possible for special states of the field conglomerate. Clearly, our next idea is to consider exact solutions for a controlled subluminal and superluminal gravity wave propagation, and we think that this program, inspired by the event GW170817/GRB 170817A, is a sufficient motivation for the presented first step.

The term {\it pp-waves} appeared in the Einstein theory of gravity, and it is usually associated with the plane-fronted gravitational waves with parallel rays \cite{Exact,MTW}. The corresponding exact solutions describe
the so-called pure gravitational radiation. Mathematically, the spacetime of the pp-wave type has to admit the existence of (at least) three Killing vectors, which form  the  ${\rm G}_3$ Abelian group of isometries; one of these three Killing vectors has to be the covariant constant null four-vector \cite{Exact}. A number of criteria are formulated, which allow us to recognize the pure gravitational radiation; one of them is that all curvature invariants have to be equal to zero \cite{Z}. The Einstein-Maxwell theory gave us the example of pure gravitational - electromagnetic radiation; the corresponding exact solution for the gravity field is of the pp-wave type, the electromagnetic field inherits the pp-wave symmetry and is characterized by vanishing quadratic invariants \cite{Exact}.
Generally, when the theory includes the cosmological constant and / or  massive scalar, pseudoscalar, vector fields, the pp-wave symmetry can not be admissible as a global symmetry of the model. However,
there is still a possibility to find specific (truncated) states of the physical system, which possesses the pp-wave symmetry. We indicate such states by the term {\it pp-wave modes} in contrast to the term {\it pure pp-waves}. In order to explain the terminological difference between pure pp-waves and pp-wave modes, we would like to make the following comment.
For instance, for the case of pp-wave symmetry the trace of the Einstein tensor in the left-hand side of the Einstein equations is equal to zero, insisting that the total stress-energy tensor in the right-hand side of these equations to be traceless also. The stress-energy tensor of the electromagnetic field in vacuum is traceless by definition; the trace of the stress-energy tensor of the massive pseudoscalar field is not equal to zero; this difference is principal: in the first case the pure pp-wave can exist without supplementary requirements, while in the second case the pp-wave mode propagation is possible only after some "fine tuning", which is  necessary, in particular, to switch off the nonvanishing trace.

A particular example of such pp-wave mode was considered recently in \cite{AB17} in the framework of the Einstein-aether theory \cite{J1,J5,J2,J6}, which is based on the introduction of global dynamic unit timelike vector field $U^i$ interpreted as a velocity four-vector of some cosmic substrate, the aether. It was shown in \cite{AB17} that the solution with pp-wave symmetry exists, when the Jacobson's constants $C_1, C_2, C_3, C_4$, introduced in the Einstein-aether theory, satisfy two requirements $C_{2}=0$ and $C_1{+}C_3=0$. Physically, this means that the aether is insensitive to shear and expansion of the aether velocity flow. In that context, the pp-wave modes became admissible, since the influence of shear and expansion, produced in the aether flow, can not be displayed because of the model restriction; as for the influence of vorticity and acceleration of the velocity flow, it is not suppressed by the requirements $C_{2}=0$, $C_1{+}C_3=0$, but is absent due to the pp-wave symmetry. In other words, if the model is truncated, the specific dynamic states, the pp-wave modes, can exist, while the model as a whole does not admit the pure pp-waves. We emphasize, that we use the term pp-wave aether modes for truncated states, keeping in mind that mathematically we deal with {\it exact solutions} to the set of equations of the Einstein-aether theory.

In this letter we consider the extended Einstein-Maxwell-aether-axion model, which describes the interaction of gravitational, electromagnetic, vector fields with a pseudoscalar field associated with axionic dark matter. Now we do not impose restrictions on the aether coupling constants $C_1,C_2,C_3,C_4$ and assume that the axionic field is massive, $m_{({\rm a})}\neq 0$. We show that now there are no pure pp-wave configurations in this model, however, there exist special solutions, which can be indicated as pp-wave aether  modes.

Why do the pp-wave aether modes exist in the model under discussion?
One can say that the pseudoscalar (axion) field is the producer of these pp-wave modes. We assume that this pseudoscalar field influences the dynamic unit vector field as a {\it switch}, suppressing "inappropriate" states of the aether flow. We deal with the set of master equations for the gravity field (supplied by electromagnetic, vector and pseudoscalar sources), for the electromagnetic field (supplied by the axionic source), for the unit vector field (supplemented by axionic guiding functions), and for the pseudoscalar field (supplied by the electromagnetic and aether sources). Surprisingly, this set of coupled master equations admits the existence of pp-wave aether modes, and the corresponding subset of exact solutions for the gravitational and electromagnetic fields coincides with the one for the pure gravitational-electromagnetic radiation.

\section{The model version of the Einstein-Maxwell-aether-axion theory}

\subsection{Action functional}

We work here with the model action functional of the following form:
$$
S = \int d^4 x \sqrt{{-}g} \left\{\frac{1}{2\kappa}\left[R +\Lambda + \lambda \left(g_{mn}U^m
U^n {-}1 \right) {+}  \right. \right.
$$
$$
\left. \left.  {+} {\cal K}^{abmn}(\phi) \nabla_a U_m  \nabla_b U_n\right] {+}  \frac14 F_{mn}F^{mn} {+}\frac14 \phi F^{mn}F^{*}_{mn} {+} \right.
$$
\begin{equation}
\left.
+ \frac{1}{2}\Psi^2_0  \left[V(\phi^2)  {-} g^{mn} \nabla_m \phi \nabla_n \phi
\right] \right\}\,.
\label{0}
\end{equation}
Here $g$ is the determinant of the metric tensor $g {=} {\rm det}(g_{ik})$; $R$ is the Ricci
scalar, $\Lambda$ is the cosmological constant; $\kappa {=} 8\pi G$, with the Newtonian coupling constant $G$ ($c{=}1$ in the chosen units); $U^i$ is the dynamic timelike vector field; $F_{mn}$ is the Maxwell tensor and $F^{*ik} \equiv \frac12 \epsilon^{ikmn}F_{mn}$ is its dual with the Levi-Civita tensor $\epsilon^{ikmn}$; $\phi$ is the pseudoscalar (axion) field with the potential $V(\phi^2)$; $\Psi_0$ is the constant reciprocal to the parameter of the axion-photon coupling $g_{A \gamma \gamma}$; $\nabla_i$ is the covariant derivative. The term $\lambda \left(g_{mn}U^m U^n {-}1 \right) $ ensures
that the $U^i$ is normalized to one; $\lambda$ is the Lagrange multiplier.
The term ${\cal K}^{abmn}(\phi^2) \ \nabla_a U_m \ \nabla_b U_n $ is quadratic in the covariant derivative
$\nabla_a U_m $ of the vector field $U^i$; the tensor ${\cal K}^{abmn}$ is constructed
using three ingredients: the metric tensor $g^{ij}$, aether velocity four-vector $U^k$, and pseudoscalar (axion) field $\phi$:
$$
{\cal K}^{abmn}(\phi^2) = h_1(\phi^2) g^{ab} g^{mn} {+} h_2(\phi^2) g^{am}g^{bn}
{+}
$$
\begin{equation}
+h_3(\phi^2)g^{an}g^{bm} {+} h_4(\phi^2) U^{a} U^{b}g^{mn}.
\label{2}
\end{equation}
The extension of this type was proposed in \cite{B16}, and now we apply this idea to the model with the pp-wave symmetry.
We assume that when $\phi=0$ the functions $h_1$, $h_2$, $h_3$, $h_4$
convert into the Jacobson's coupling constants $C_1$, $C_2$, $C_3$ and $C_4$   \cite{J1}:
\begin{equation}
h_1(0)= C_1 \,, h_2(0)= C_2 \,, h_3(0)= C_3 \,, h_4(0)= C_4 \,.
\label{23}
\end{equation}
 All admissible terms in the Lagrangian, which include convolutions of the velocity four-vector $U^i$ and its covariant derivative $\nabla_mU_n$ with the Maxwell tensor $F_{mn}$ (in the linear and quadratic forms), are classified and discussed in \cite{AE10}; here we consider the classical terms $\frac14 F_{mn}F^{mn}$  and $\frac14 \phi F^{mn}F^{*}_{mn}$ only.

\subsection{Potential of the axion field}

We assume that the potential of the pseudoscalar (axion) field is bi-quadratic with respect to $\phi$:
\begin{equation}
V(\phi^2) = \frac12  \left[\frac{m^2_{({\rm a})}}{\nu} + \nu (\phi^2-\phi^2_{0})\right]^2  \,.
\label{b1}
\end{equation}
Here $m_{({\rm a})}$ is the parameter associated with an intrinsic axion mass; $\nu$ is the parameter associated with the so-called self-interaction of the $\phi^4$ type; $\phi_0$ relates to some basic state of the pseudoscalar (axion) field. This potential possesses the following properties. First, when
\begin{equation}
\phi = \phi_{*} \equiv \pm \sqrt{\phi^2_{0} - \frac{m^2_{({\rm a})}}{\nu^2}} \,,
\label{b19}
\end{equation}
the potential and its derivative take zero values:
\begin{equation}
V(\phi_{*}^2) = 0 \,, \quad
\left[\frac{d}{d\phi}V(\phi_{}^2)\right]_{|\phi=\phi_{*}} = 0 \,.
\label{b11}
\end{equation}
Second, when $|\phi_{0}| > \frac{m_{({\rm a})}}{|\nu|}$, this potential
has two symmetric minima coinciding with double zeros of this
function, and one maximum at $\phi=0$. Third, when $\phi_0=0$, $|\phi|$ is small, and $\frac{m_{({\rm a})}}{|\nu|}$ is absorbed by the cosmological constant, the potential tends to the standard quantity $V \to m^2_{({\rm a})} \phi^2$, which describes massive pseudoscalar field without self-interaction.

\subsection{Decomposition of the covariant derivative of the aether velocity four-vector}

The tensor $\nabla_i U_k$ can be decomposed into a sum of irreducible parts, the acceleration four-vector $DU^{i}$,
the shear tensor $\sigma_{ik}$, the vorticity tensor $\omega_{ik}$, and
the expansion scalar $\Theta$, as follows:
\begin{equation}
\nabla_i U_k = U_i DU_k + \sigma_{ik} + \omega_{ik} +
\frac{1}{3} \Delta_{ik} \Theta \,. \label{act3}
\end{equation}
The basic quantities in this decomposition are defined as
$$
DU_k \equiv  U^m \nabla_m U_k \,, \quad \sigma_{ik}
\equiv \frac{1}{2}\left(\nablab_i U_k {+}
\nablab_k U_i \right) {-} \frac{1}{3}\Delta_{ik} \Theta  \,,
$$
$$
\omega_{ik} \equiv \frac{1}{2} \left(\nablab_i U_k {-} \nablab_k U_i \right) \,, \quad \Theta \equiv \nabla_m U^m
\,,
$$
\begin{equation}
D \equiv U^i \nabla_i \,, \quad \Delta^i_k = \delta^i_k - U^iU_k \,, \quad \nablab_i \equiv \Delta_i^k \nabla_k \,. \label{act4}
\end{equation}
In these terms the scalar ${\cal K}^{abmn}(\nabla_a U_m) (\nabla_b U_n)$ takes the form
$$
{\cal K}^{abmn}(\nabla_a U_m) (\nabla_b U_n) =
$$
$$
=(h_1 {+} h_4)DU_k DU^k {+}
(h_1 {+} h_3)\sigma_{ik} \sigma^{ik} {+}
$$
\begin{equation}
+ (h_1 {-} h_3)\omega_{ik}
\omega^{ik} {+} \frac13 \left(h_1 {+} 3h_2 {+}h_3 \right) \Theta^2
\,, \label{act5}
\end{equation}
and gives us the idea to make the following redefinitions
$$
h_{D}(\phi^2) \equiv h_1+h_4 \,, \quad
h_{\sigma}(\phi^2) \equiv h_1+h_3 \,,
$$
\begin{equation}
h_{\omega}(\phi^2) \equiv h_1-h_3 \,, \quad  h_{\theta}(\phi^2) \equiv \frac13 (h_1+h_3) + h_2 \,.
\label{w120}
\end{equation}
Below we indicate the quantities $h_{D}(\phi^2)$, $h_{\sigma}(\phi^2)$, $h_{\omega}(\phi^2)$, $h_{\theta}(\phi^2)$ as guiding functions.

\subsection{Ansatz about the structure of guiding functions}

We assume the guiding functions to have the form of {\it switch-functions}:
\begin{equation}
h_{D}(\phi^2) = (C_1+C_4) \left(1- \frac{\phi^2}{\phi^2_{D}} \right)^{n_D} \,,
\label{w13}
\end{equation}
\begin{equation}
h_{\sigma}(\phi^2) = (C_1+C_3) \left(1- \frac{\phi^2}{\phi^2_{\sigma}} \right)^{n_{\sigma}}\,,
\label{w11}
\end{equation}
\begin{equation}
h_{\omega}(\phi^2) = (C_1-C_3) \left(1- \frac{\phi^2}{\phi^2_{\omega}} \right)^{n_{\omega}} \,,
\label{w12}
\end{equation}
\begin{equation}
h_{\theta}(\phi^2) =
 \left[\frac13 (C_1{+}C_3){+}C_2\right] \left(1-\frac{\phi^2}{\phi^2_{\theta}} \right)^{n_{\theta}} \,,
\label{w14}
\end{equation}
where the integer indices $n_D$, $n_{\sigma}$, $n_{\omega}$ and $n_{\theta}$ are the numbers exceeding one,  $n_D>1$, $n_{\sigma}>1$, $n_{\omega}>1$ and $n_{\theta}>1$.
These inequalities guarantee that
\begin{equation}
h_{s}(\phi^2_{s}) = 0 \,, \quad  h^{\prime}_{s}(\phi^2_{s}) = 0 \,,
 \label{w911}
\end{equation}
for $s=D,\sigma,\omega,\theta$, thus explaining the terms {\it switch on} and {\it switch off}, which we use below. The guiding functions coincide with the corresponding combinations of the Jacobson's constants in two cases: first, when $\phi \equiv 0$ (absence of the axion field); second, in the limit of large critical values $\phi^2_{s} \to \infty$ (axion influence is suppressed). When $\phi^2$ grows and overcomes the critical value $\phi^2_{s}$, some of guiding functions can change the signs, if the corresponding integer indices are odd; the guiding functions hold the signs if the indices are even. Here we do not fix these indices.

\subsection{About critical values of the axion field}

The presented model contains five critical parameters, which characterize the state of the pseudoscalar (axion) field: $\phi_{*}$, $\phi_{D}$, $\phi_{\sigma}$, $\phi_{\omega}$, $\phi_{\theta}$, and four integer numbers $n_{D}$, $n_{\sigma}$, $n_{\omega}$, $n_{\theta}$. There is a unique case, when all five critical parameters coincide, and all numbers also coincide. Another interesting case is when four switch parameters coincide, but they differ from the critical parameter $\phi_{*}$:
\begin{equation}
\phi_{D} = \phi_{\sigma} = \phi_{\omega} = \phi_{\theta} \equiv \phi^{**} \,, \quad \phi_{*} \neq \phi^{**} \,.
\label{sw1}
\end{equation}
But the most interesting (from our point of view) are the cases, when the parameter $\phi_{*}$ coincides with one or two switch parameters. In particular, below we consider the case, when $\phi_{*}=\phi_{\sigma}=\phi_{\theta}$, and show that this condition is necessary for existence of the pp-wave aether modes.

\subsection{Master equation for the axion field}

Variation of the action functional (\ref{0}) with respect to the pseudoscalar field $\phi$ gives the master equation
\begin{equation}
\nabla^m \nabla_m \phi + \phi \left[V^{\prime}(\phi^2) + \frac{{\cal H}}{\kappa \Psi^2_0}\right] = - \frac{1}{4\Psi^2_0} F^{*}_{mn}F^{mn}  \,,
\label{ax1}
\end{equation}
where the regulator function ${\cal H}$:
$$
{\cal H} = h^{\prime}_{D}(\phi^2) \ DU_m DU^m  {+} h^{\prime}_{\sigma}(\phi^2) \ \sigma_{ik} \sigma^{ik} {+}
$$
\begin{equation}
{+} h^{\prime}_{\omega}(\phi^2) \ \omega_{ik} \omega^{ik} {+} h^{\prime}_{\theta}(\phi^2) \ \Theta^2 \,,
\label{ax3}
\end{equation}
contains the first derivatives of the guiding functions in front of the squares of acceleration four-vector, shear tensor, vorticity tensor and expansion scalar, respectively.

\subsection{Master equations for the electromagnetic field}

Variation of the action functional (\ref{0}) with respect to $A_i$, the potential of the electromagnetic field, yields the equations of axion electrodynamics
\begin{equation}
\nabla_k \left[F^{ik} + \phi F^{*ik} \right] = 0 \,.
\label{M1}
\end{equation}
In addition, we use the equation
\begin{equation}
\nabla_k F^{*ik} = 0 \,,
\label{M2}
\end{equation}
which is the direct consequence of the definition of the Maxwell tensor $F_{mn} = \nabla_m A_n {-} \nabla_n A_m$. As usual, we assume that $A_k$ satisfies the Lorentz gauge, $\nabla_k A^k=0$. The most known form of equations (\ref{M1}) is
 \begin{equation}
\nabla_k F^{ik} = - \frac12 \epsilon^{ikmn}F_{mn} \nabla_k \phi \,,
\label{M3}
\end{equation}
which uses directly the equation (\ref{M2}) and the definition of the dual Maxwell tensor. The equations (\ref{M3}), (\ref{M2}) and (\ref{ax1}) with ${\cal H}=0$ form the set of basic equations for standard axion electrodynamics (see, e.g., \cite{AE1,AE2,AE3,AE4,AE5,AE6,AE7,AE8,AE9} for details, history  and applications). In our case ${\cal H}\neq 0$, thus we deal with one of the variants of extension of the axion electrodynamics \cite{symmetry}.

\subsection{Master equations for the unit vector field}

The variation of the action (\ref{0}) with respect to
$\lambda$ yields the equation
\begin{equation}
g_{mn}U^m U^n = 1 \,,
\label{21}
\end{equation}
which is known to be the normalization condition of the timelike
vector field $U^k$.
Then, variation of the functional (\ref{0}) with respect to
$U^i$ yields that $U^i$ itself satisfies the well-known equations
\cite{J1}
\begin{equation}
\nabla_a {\cal J}^{aj} = I^j + \lambda \ U^j  \,.
\label{0A1}
\end{equation}
Here the following definitions are used:
\begin{equation}
{\cal J}^{aj} = {\cal K}^{abjn}(\phi^2) \ \nabla_b U_n  \,,
\label{J2}
\end{equation}
\begin{equation}
I^j =  h_4(\phi^2) \ DU_m \nabla^j U^m  \,.
\label{J6}
\end{equation}
The Lagrange multiplier
\begin{equation}
\lambda =  U_j \left[\nabla_a {\cal J}^{aj}- I^j \right]   \label{0A309}
\end{equation}
is obtained by convolution of (\ref{0A1}) with $U_j$.

\subsection{Master equations for the gravitational field}

The variation of the action (\ref{0}) with respect to the metric
$g^{ik}$ yields the gravitational field equations, which can be presented in the following form:
$$
R_{ik} - \frac{1}{2} R \ g_{ik} - \Lambda g_{ik} =
$$
\begin{equation}
=\lambda U_i U_k  + T^{({\rm U})}_{ik} +
\kappa T^{({\rm A})}_{ik} + \kappa T^{({\rm M})}_{ik}
\,. \label{0Ein1}
\end{equation}
The term $T^{({\rm U})}_{ik}$ describes the contribution of
the vector field $U^i$ into the total stress-energy tensor:
$$
T^{({\rm U})}_{ik} =
\frac12 g_{ik} {\cal J}^{am} \nabla_a U_m {+}
$$
$$
{+}\nabla^m \left[U_{(i}{\cal J}_{k)m}\right] {-}
\nabla^m \left[{\cal J}_{m(i} U_{k)} \right] {-}
\nabla_m \left[{\cal J}_{(ik)} U^m\right]+
$$
$$
+h_1\left[(\nabla_mU_i)(\nabla^m U_k) {-}
(\nabla_i U_m \nabla_k U^m) \right] {+}
$$
\begin{equation}
{+}h_4 (U^a \nabla_a U_i)(U^b \nabla_b U_k) \,.
\label{5Ein1}
\end{equation}
As usual, the symbol $p_{(i} q_{k)}{\equiv}\frac12 (p_iq_k{+}p_kq_i)$
denotes the procedure of symmetrization. The tensor
\begin{equation}
T^{({\rm A})}_{ik} = \Psi^2_0 \left\{\nabla_i \phi \nabla_k \phi + \frac12 g_{ik}\left[V(\phi^2) {-} \nabla_n \phi \nabla^n \phi \right]\right\}
\label{7Ein1}
\end{equation}
describes the contribution of the axion field into the total stress-energy tensor.
The last term
\begin{equation}
T^{({\rm M})}_{ik} = \frac14 g_{ik} F_{mn} F^{mn} - F_{im} F_k^{\ \ m} \,, \label{TM}
\end{equation}
is the canonic stress-energy tensor of the electromagnetic field.
Compatibility conditions for the set of equations (\ref{0Ein1})
\begin{equation}
\nabla^k\left[ \lambda U_i U_k  + T^{({\rm U})}_{ik} + \kappa T^{({\rm A})}_{ik} + \kappa T^{({\rm M})}_{ik}
\right] = 0
 \label{compa1}
\end{equation}
are satisfied automatically on the solutions to master equations (\ref{ax1})-(\ref{0A1}).

\section{PP-wave aether modes }

We consider below the model with the so-called pp-wave symmetry; the symmetry of this model is associated with two aspects: the geometrical aspect (the spacetime isometries), and the physical aspect (symmetry inheritance by electromagnetic, vector and pseudoscalar fields).

\subsection{Geometry of the pp-wave state}

The Lie derivative of the metric is equal to zero,
$\pounds_{\xi^l_{(\alpha)}} g_{ik} =0$, along the triplet of Killing
vectors $\{\xi^i_{(1)}, \xi^i_{(2)}, \xi^i_{(3)} \}$, which form
the Abelian group of isometries ${\rm G}_3$; $\xi^i_{(1)}$ is
the null covariant constant Killing vector, i.e.,
$g_{ik}\xi^i_{(1)}\xi^k_{(1)}=0$, and $\nabla_k \xi^i_{(1)}=0$.
The metric describing the gravitational pp-wave of the first polarization can be chosen in the form (see, e.g., \cite{MTW} for details)
\begin{equation}
ds^2 = 2 du dv - L^2 \left(e^{2\beta} {dx^2}^2 +
e^{-2\beta} {dx^3}^2
\right) \,.
\label{PW1}
\end{equation}
Here $u$ and $v$ are the retarded and advanced times,
respectively, given in terms of the time $t$ and spatial
coordinate $x^1$ by $u{=}\frac{1}{\sqrt2}(t{-}x^1)$,
$v{=}\frac{1}{\sqrt2}(t{+}x^1)$, and $x^2, x^3$ are the spatial
coordinates in the plane of the front of the pp-wave aether mode. The
quantities $L(u)$ and $\beta(u)$ are  functions of the retarded
time $u$ only. The three Killing vectors in this case are known to
be of the form
\begin{equation}
\xi^i_{(1)} = \delta^i_v \,, \quad  \xi^i_{(2)} = \delta^i_2 \,, \quad \xi^i_{(3)} = \delta^i_3 \,,
\label{K1}
\end{equation}
the first of them is the null four-vector, i.e.,
$g_{ik}\xi^i_{(1)} \xi^k_{(1)} =0$; also the four-vectors are
orthogonal one to another.

For the spacetime with such symmetry the Riemann tensor is characterized by two nonvanishing components $R^2_{\ u2u}$ and $R^3_{\ u3u}$; the Ricci tensor has only one nonvanishing component $R_{uu}$; the Ricci scalar is equal to zero. As in the classical case with pure pp-waves, we assume that $\Lambda {=}0$.  Also, we have to take into account that the total stress-energy tensor in the right-hand side of (\ref{0Ein1}) should be traceless (since $R=0$), and orthogonal to all three Killing vectors (since $R_{ik}\xi^k_{(\alpha)}=0$).

\noindent
{\it REMARK}

\noindent
 Let us emphasize, that we use the metric (\ref{PW1}) in the form, which is indicated in the book \cite{MTW} as the metric in the TT-gauge. There is an alternative representation of the metric with pp-wave symmetry in the so-called Fermi coordinates, described by the function $H(u,y,z)$ (see, e.g., \cite{Exact} for mathematical details and references). One can recall, for instance, that
the metric
$$
\mbox{d}s^2 = 2 \mbox{d} \bar{u} \mbox{d}\bar{v}- \mbox{d}y^2
-\mbox{d}z^2 - 2 {\cal H} (\bar{u},y,z) \mbox{d}\bar{u}^2 \ ,
$$
\begin{equation}
\bar{u} = \frac{t - x}{\sqrt{2}}\ , \quad  \bar{v} = \frac{t + x}{\sqrt{2}}
\label{27}
\end{equation}
with the harmonic function
\begin{equation}
2{\cal H}(\bar{u},y,z) = A(\bar{u}) \left( y^2 - z^2 \right) + 2B(\bar{u}) \ y z
\label{29}
\end{equation}
corresponds to (\ref{PW1}), if we use the transformations
\begin{eqnarray}
\bar{u} &=& u , \quad  \bar{v} = v + \frac{1}{4}
\left[(x^2)^2 (L^2 e^{2\beta})^{\displaystyle \cdot}
+  (x^3)^2(L^2 e^{-2\beta})^{\displaystyle \cdot} \right]\ ,
\nonumber\\
y &=& L e^{\beta} \cdot  x^2 , \quad z = L e^{-\beta} \cdot  x^3  ,
\quad
A(u) = \ddot{\beta} {+} 2 \dot{\beta} \frac{\dot{L}}{L} .
\label{34}
\end{eqnarray}
(In this Remark the dot indicates the differentiation with respect to the retarded time $\bar{u} {=} u$).

\subsection{Symmetry inheritance}

We assume that the vector, electromagnetic  and pseudoscalar fields inherit the
pp-wave symmetry; mathematically, this means that
\begin{equation}
\pounds_{\xi^i_{(\alpha)}} U^k =0 \,, \quad \pounds_{\xi^i_{(\alpha)}} A_j =0 \,, \quad
\pounds_{\xi^i_{(\alpha)}} \phi =0
\label{sym1}
\end{equation}
for all three Killing vectors (\ref{K1}). These requirements provide that the vector, electromagnetic  and pseudoscalar fields do not depend on $v,
x^2,x^3$, and thus are the functions of the retarded time only: $U^i(u)$, $A_k(u)$, $\phi(u)$.

\subsubsection{The consequence for the electromagnetic field structure}

The pp-wave symmetry inheritance requires the potential $A_k$ to contain only two nonvanishing components $A_2(u)$ and $A_3(u)$, which are arbitrary functions of the retarded time. Indeed,
the Lorentz gauge leads to the relationship $\left[L^2(u) A_v(u)\right]^{\prime}{=}0$, or $A_v(u){=}\frac{const}{L^2}$; for physical reasons we assume this constant is equal to zero. (Here and below the prime denotes the derivative with respect to the retarded time $u$). The component $A_u(u)$ does not contribute to the corresponding Maxwell tensor, which has now only two components $F_{u2}$ and $F_{u3}$. Clearly, for this structure of the Maxwell tensor the equations of axion electrodynamics (\ref{M3}) are satisfied identically for arbitrary $A_2(u)$, $A_{3}(u)$, since $\nabla_k \phi = \delta_k^u \phi^{\prime}(u)$. Finally, one can see directly, that both invariants of the electromagnetic field vanish, $F_{mn}F^{mn}{=}0$, $F^{*}_{mn}F^{mn}{=}0$, thus we deal with the so-called pure electromagnetic radiation.

\subsubsection{The consequence for the unit vector field structure}

Since the requirement $\pounds_{\xi^i_{(\alpha)}} U^k =0$ with the
Killing vectors (\ref{K1}) guaranties that the components of the four-vector $U^i$ can be
the functions of the retarded time only, we focus on the simplest case
\begin{equation}
U^i = \frac{1}{\sqrt2} \left(\delta^i_u + \delta^i_v \right) \,,
\label{K2}
\end{equation}
which describes the aether in the state of rest in the chosen frame of reference and metric (\ref{PW1}).
The covariant derivative of the velocity four-vector
reduces as follows:
\begin{equation}
\nabla_i U^k  = \frac{1}{\sqrt2} \left[\delta_i^2
\delta^k_2 \left(\frac{L^{\prime}}{L}{+}
\beta^{\prime} \right) {+} \delta_i^3 \delta^k_3
\left(\frac{L^{\prime}}{L}{-} \beta^{\prime} \right)\right]\,,
\label{GW001}
\end{equation}
providing the acceleration four-vector and the
vorticity tensor to be vanishing
\begin{equation}
DU^k = 0 \,, \quad \omega_{pq} = 0
\,.
\label{PW3bef}
\end{equation}
The expansion scalar is proportional to the derivative of the so-called background factor $L$:
\begin{equation}
\Theta = \frac{\sqrt{2}\,L^{\prime}(u)}{L} \,.
\label{GW0011}
\end{equation}
The shear tensor is also nonvanishing:
\begin{equation}
\sigma^k_i  = \frac{\Theta}{2}  \left(\frac13
\Delta_i^k - \delta_i^1 \delta^k_1 \right)
+ \frac{\beta^{\prime}}{\sqrt2} \left(\delta_i^2 \delta^k_2
{-} \delta_i^3 \delta^k_3 \right)\,.
\label{GW002}
\end{equation}

\subsubsection{The consequence for the axion field structure}

For the axion field, which depends on the retarded time only, $\phi(u)$, we obtain that
\begin{equation}
\nabla^m \nabla_m \phi = g^{mn} \left[\partial_m \partial_n \phi - \Gamma^s_{mn}\partial_s \phi \right] = 0 \,,
\label{ax11d}
\end{equation}
since $g^{uu}{=}0$, and $\Gamma^u_{mn} {=}0$. Also, the second (pseudo)invariant of the electromagnetic field vanishes, $F^{*}_{mn}F^{mn}=0$, thus, the equation (\ref{ax1}) reduces to
\begin{equation}
\phi \left\{V^{\prime}(\phi^2) {+} \frac{{1}}{\kappa \Psi^2_0} \left[h^{\prime}_{\sigma}(\phi^2) \ \sigma_{ik} \sigma^{ik} {+}  h^{\prime}_{\theta}(\phi^2) \ \Theta^2
 \right]\right\} {=} 0  \,.
\label{ax11}
\end{equation}
One of solutions to this equation is trivial, $\phi {=} 0$. To guarantee that the term in braces vanishes for arbitrary scalars $\sigma_{ik} \sigma^{ik}$ and $\Theta^2$, we have to require, that
for chosen solutions $V^{\prime}(\phi^2){=0}$, $h^{\prime}_{\sigma}(\phi^2){=}0$ and $h^{\prime}_{\theta}(\phi^2){=}0$. It is possible, when
\begin{equation}
\phi^2_{*} = \phi^2_{\sigma} = \phi^2_{\theta} \,, \quad \phi = \pm \phi_{*} \,.
\label{ax11b}
\end{equation}
The admissible solution of the pp-wave type happens to be constant; it delivers the minimum to the axion potential and turns to zero the axion potential $V$ and two guiding functions $h_{\sigma}$ and $h_{\theta}$.

\subsection{Exact solutions}

In order to describe exact solutions indicated as pp-wave aether modes, we have to solve the master equations for the gravitational, electromagnetic, axion and vector fields.
The solution to the equations of axion electrodynamics is already found, $A_i {=} \delta_i^2 A_2(u){+}\delta_i^3 A_3(u)$. The solution to the axion field equation is constant, $\phi {=} \pm \phi_{*} = \pm \phi_{\sigma} = \pm \phi_{\theta}$. Since the acceleration four-vector and vorticity tensor are absent, the dynamic tensor ${\cal J}^{mn}$, obtained for the case
$h_{\sigma}(\phi^2_{*}){=}0$ and $h_{\theta}(\phi^2_{*}){=}0$ is equal to zero, ${\cal J}^{am} = 0$. Similarly, we obtain that
$I^n(u) = 0$ and thus $\lambda=0$. In other words, the master equations for the unit vector field is satisfied identically.

The master equations for the gravitational field (see (\ref{0Ein1}) with $\Lambda=0$) reduce now to the one key equation:
\begin{equation}
L^{\prime \prime}(u) {+} L\left\{{\beta^{\prime}}^2  {+} \frac{\kappa}{2L^2}  \left[\left(A^{\prime}_2 e^{{-}\beta}\right)^2 {+} \left(A^{\prime}_3
e^{\beta}\right)^2 \right]\right\} =0 \,. \label{Lpp}
\end{equation}
Indeed, the stress-energy tensor $T^{({\rm U})}_{ik}$ is equal to zero, since ${\cal J}^{mn}=0$; the tensor $T^{({\rm A})}_{ik}$ is equal to zero, since $\nabla_k \phi_{*}=0$ and $V(\phi^2_{*})=0$;
the stress-energy tensor $T^{({\rm M})}_{ik}$ contains only the contribution $T^{({\rm M})}_{uu} = {-}(F_{u2}F_u^{\ 2}{+} F_{u3}F_u^{\ 3})$.
The functions $\beta(u)$, $A_2(u)$ and $A_3(u)$ are arbitrary. In order to clarify physical sense of terms in the equation (\ref{Lpp}), we use the so-called tetrad components of the electric field, based on the following relationships:
$$
E_2 = F_{2k} U^k = -\frac{A^{\prime}_2(u)}{\sqrt2}  \,, \quad E_3 = F_{3k} U^k = -\frac{A^{\prime}_3(u)}{\sqrt2}  \,,
$$
\begin{equation}
{\cal E}_{(2)} \equiv \sqrt{{-} E_2 E^2} = \frac{A^{\prime}_2 e^{-\beta}}{\sqrt2 L}  \,, \quad
{\cal E}_{(3)} \equiv \sqrt{{-} E_3 E^3} = \frac{A^{\prime}_3 e^{\beta}}{\sqrt2 L} \,. \label{tetra1}
\end{equation}
In these terms the equation (\ref{Lpp}) takes the form
\begin{equation}
L^{\prime \prime}(u) {+} L\left\{{\beta^{\prime}}^2  {+} \kappa \left[{\cal E}^2_{(2)} + {\cal E}^2_{(3)} \right]\right\} =0 \,. \label{Lpp2}
\end{equation}
The equation (\ref{Lpp2}) is the differential equation of the second order in ordinary derivatives for the unknown function $L(u)$ and three arbitrary functions $\beta(u)$, ${\cal E}_{(2)}(u)$, ${\cal E}_{(3)}(u)$. In order to add the initial data to this equation, we indicate the front of the pp-wave aether mode by the condition $u=0$, and require that on this front
\begin{equation}
L(0) = 1 \,, \quad L^{\prime}(0)=0 \,.  \label{act59}
\end{equation}
This equation is well-known, and it was studied in many works. As an example, we discuss the case, when
$$
{\cal E}_{(2)}(u) = E_0 \cos{\omega u} \cos{\Omega u} \,,
$$
$$
{\cal E}_{(3)}(u) = E_0 \cos{\omega u} \sin{\Omega u}  \,,
$$
\begin{equation}
\beta(u) = \beta_0 (1-\cos{\omega u}) \,, \quad \beta(0)=0 \,, \quad \beta^{\prime}(0) =0 \,.  \label{L44}
\end{equation}
Then the equation (\ref{Lpp2}) transforms into the well-known Mathieu equation
\begin{equation}
L^{\prime \prime}(u) {+} L\left[\Pi^2 - h \cos{2\omega u} \right] =0 \,,  \label{L46}
\end{equation}
where we use the definitions
\begin{equation}
\Pi^2 = \frac12 \left(\beta^2_0 \omega^2 +\kappa E^2_0 \right) \,, \quad h = \frac12 \left(\beta^2_0 \omega^2 - \kappa E^2_0 \right) \,.  \label{L45}
\end{equation}
The behavior of the Mathieu functions is well documented. For illustration, we consider the explicit particular solution for the case $h=0$, i.e., $\beta_0 \omega = \sqrt{\kappa} E_0$:
\begin{equation}
L(u) = \cos{\Pi_0 u} \,, \quad \Pi_0 = \sqrt{\kappa} E_0\,.  \label{L450}
\end{equation}
Clearly, the background factor $L(u) = (-g)^{\frac14}$ starts from one at $u=0$ and reaches zero at $u=\frac{\pi}{2\Pi_0}$; thus, this  solution is well defined on the interval $0 \leq u<\frac{\pi}{2\Pi_0}$.

\section{Conclusions}

In the framework of the extended Einstein-Maxwell-aether-axion model, we have found new exact solutions to the system of coupled evolutionary equations, which possesses pp-wave symmetry. We indicated these solutions as the pp-wave aether modes. It seems to be interesting to distinguish the following three remarks concerning the obtained exact solutions.

\noindent
1. There are four players in the scheme of internal interactions in the aether model under consideration: the gravitational, electromagnetic, timelike unit vector, and pseudoscalar (axion) fields. The axion field is the key player in this quartet. It contributes in a standard manner the gravity field equations (see the source (\ref{7Ein1}) in (\ref{0Ein1})); its gradient four-vector forms the current-like term in the equations of axion electrodynamics (\ref{M3}). But the main new feature in the  {\it modus operandi} of the axion field is the control of the aether behavior through the guiding functions (\ref{w13})-(\ref{w14}) incorporated into the constitutive tensor (\ref{2}). In this sense the aether can be called axionically active.

\noindent
2. The axionic guiding functions (\ref{w13})-(\ref{w14}) convert into Jacobson's coupling constants $C_1$, $C_2$, $C_3$, $C_4$ (arbitrary parameters in this approach), when the pseudoscalar field takes zero value, corresponding to the local maximum of the axionic potential (\ref{b1}). Clearly, an arbitrary constant is not a solution for the axion field.
However, when the axion field takes one of the critical values (\ref{b19}), corresponding to one of the local minima of the potential (\ref{b1}), the following new possibility appears: one, two, three or four guiding functions can vanish, if the minimal value (\ref{b19}) coincides, respectively, with one, two, three or four critical values incorporated into (\ref{w13})-(\ref{w14}). From the physical point of view, this effect looks like switch-on / switch-off procedure: depending on the chosen critical value, the axion field can switch off the influence
of the acceleration, shear, vorticity, expansion of the aether velocity flow.
The pp-wave aether modes can appear in the system, when the axion field switches off the influence of shear and expansion simultaneously. Respectively, one can consider the appearance of pure rotational aether modes, pure shear aether modes, etc., if the axion field can switch off other types of influence of the velocity field flow; we intend to consider these effects in the nearest future.

\noindent
3. The aether pp-wave modes described in this letter, in fact, do not activate scalar mode (spin-0 mode in terminology of the paper \cite{J5}), and the tensor (spin-2) mode propagates with the velocity equal to the speed of light in vacuum. These results are obtained in \cite{AB17} using special requirements for the Jacobson's coupling constants, but now it is the result of axionic regulation of the aether dynamics. The next problem is the stability of the solutions of the pp-wave type, but this task is beyond the scope of this letter. Finally, one can imagine a number of mechanisms, which could switch off the pp-wave aether modes. For instance, let the perturbation of the electromagnetic field transform the pure electromagnetic radiation into the waves with nonvanishing second quadratic invariant; now the equation (\ref{ax1}) with nonvanishing right-hand side does not admit constant solution; as a consequence, the guiding functions (\ref{w11}) and (\ref{w14}) are not equal to zero, thus stopping the propagation of the aether pp-wave modes. In other words, one can destroy the pp-wave aether modes by introduction of the electromagnetic perturbations; as for the inverse process, the excitation of the pp-wave aether modes, we hope to consider the corresponding processes in the next paper.

\acknowledgments{The work was supported by Russian Science Foundation (Project No. 16-12-10401), and, partially, by the Program of Competitive Growth
of Kazan Federal University.}

\end{document}